\documentclass[%
  reprint,
  showpacs,preprintnumbers,
  amsmath,amssymb,
  aps,
  prb,
]{revtex4-1}

\usepackage{graphicx}
\usepackage{dcolumn}
\usepackage{bm}
\usepackage{hyperref}
\usepackage{multirow}
\usepackage{color}
\usepackage{ulem}

\usepackage{times}
\usepackage{booktabs}

\begin{document}

\preprint{APS/123-QED}

\title{Emergent odd-parity multipoles and magnetoelectric effects on a diamond structure:\\ implication to 5$d$ transition metal oxides $A$OsO$_4$ ($A=$ K, Rb, and Cs)}

\author{Satoru Hayami,$^1$ Hiroaki Kusunose,$^2$ and Yukitoshi Motome$^3$}
\affiliation{%
 $^1$Department of Physics, Hokkaido University, Sapporo 060-0810, Japan \\
 $^2$Department of Physics, Meiji University, Kawasaki 214-8571, Japan \\
 $^3$Department of Applied Physics, University of Tokyo, Tokyo 113-8656, Japan
}%
 
\begin{abstract}
We report our theoretical predictions on the linear magnetoelectric (ME) effects originating from odd-parity multipoles associated with spontaneous spin and orbital ordering on a diamond structure. 
We derive a two-orbital model for $d$ electrons in $e_g$ orbitals by including the effective spin-orbit coupling which arises from the mixing between $e_g$ and $t_{2g}$ orbitals. 
We show that the model acquires a net antisymmetric spin-orbit coupling once staggered spin and orbital orders occur spontaneously. 
The staggered orders are accompanied by odd-parity multipoles: magnetic monopole, quadrupoles, and toroidal dipoles. 
We classify the types of the odd-parity multipoles according to the symmetry of the spin and orbital orders. 
Furthermore, by computing the ME tensor using the linear response theory, we show that the staggered orders induce a variety of the linear ME responses. 
We elaborate all possible ME responses for each staggered order, which are useful to identify the order parameter and to detect the odd-parity multipoles by measuring the ME effects. 
We also elucidate the effect of lowering symmetry by a tetragonal distortion, which leads to richer ME responses. 
The implications of our results are discussed for $5d$ transition metal oxides, $A$OsO$_4$ ($A=$ K, Rb, and Cs), in which the order parameters are not fully identified. 

\end{abstract}
\maketitle

\section{Introduction}
\label{sec:Introduction}
The effect of the relativistic spin-orbit coupling (SOC) in solids has drawn considerable interest in condensed matter physics. 
In particular, the SOC in the absence of spatial inversion symmetry, which is called the antisymmetric spin-orbit coupling (ASOC), has widely been studied as a source of intriguing phenomena in materials with noncentrosymmetric lattice structures, e.g., an unconventional superconductivity in a heavy-fermion compound CePt$_3$Si~\cite{Bauer_PhysRevLett.92.027003}, giant spin splitting in the electronic band structure in a semiconductor BiTeI~\cite{ishizaka2011giant},  and the quantum spin Hall effect in transition metal dichalcogenides~\cite{qian2014quantum}.  
Once such noncentrosymmetric systems undergo the breaking of time-reversal symmetry by a magnetic order, further unusual phenomena may arise, such as magnetoelectric (ME) effects in multiferroic materials~\cite{Fiebig0022-3727-38-8-R01,KhomskiiPhysics.2.20}, the topological Hall effect in skyrmion crystals~\cite{nagaosa2013topological}, and nonreciprocal optical phenomena in Rashba metals~\cite{Kawaguchi_PhysRevB.94.235148}. 

Meanwhile, such interesting physics may occur even in centrosymmetric systems, once the lattice site lacks the inversion center~\cite{sakhnenko2012magnetoelectric,zhang2014hidden,Yanase_JPSJ.83.014703,Hayami_PhysRevB.90.024432,riley2014direct}. 
Such asymmetry at the lattice site is ubiquitously found in the centrosymmetric systems whose unit cell includes sublattices, e.g., zigzag chain, honeycomb, and diamond structures. 
In these systems, a spontaneous staggered order, such as a N\'eel-type antiferromagnetic order, breaks spatial inversion symmetry, which activates a net ASOC. 
This mechanism leads to intriguing phenomena even in centrosymmetric systems, such as the ME effect in Cr$_2$O$_3$~\cite{dzyaloshinskii1960magneto,astrov1960magnetoelectric,popov1999magnetic} and Co$_4$Nb$_2$O$_9$~\cite{fischer1972new,Khanh_PhysRevB.93.075117,Khanh_PhysRevB.96.094434,Yanagi2017}, valley splitting in the electronic band structure~\cite{li2013coupling}, nonreciprocal magnon excitations in $\alpha$-Cu$_2$V$_2$O$_7$~\cite{Gitgeatpong_PhysRevB.92.024423,Hayami_doi:10.7566/JPSJ.85.053705,Gitgeatpong_PhysRevB.95.245119,Gitgeatpong_PhysRevLett.119.047201}, and unconventional superconductivity~\cite{Sumita_PhysRevB.93.224507}. 

Behind the intriguing phenomena in centrosymmetric systems, multipoles with odd parity play an important role~\cite{hayami2016emergent}. 
The staggered electronic orders are accompanied by the odd-parity multipoles, such as magnetic quadrupole~\cite{Watanabe_PhysRevB.96.064432}, electric octupole~\cite{hitomi2014electric}, and magnetic toroidal dipoles~\cite{Spaldin_0953-8984-20-43-434203,kopaev2009toroidal,Yanase_JPSJ.83.014703,Hayami_PhysRevB.90.024432,Hayami_PhysRevB.90.081115,Hayami_doi:10.7566/JPSJ.84.064717,Hayami_1742-6596-592-1-012101,Hayami_doi:10.7566/JPSJ.85.053705,yanagi2017optical}. 
A microscopic theory has recently been elucidated by the authors through the systematic analysis of staggered charge, spin, and orbital orderings in a minimal model on a honeycomb structure~\cite{Hayami_PhysRevB.90.081115, hayami2016emergent,hayami2017mean}. 
The authors constructed a classification of the staggered orderings and associated odd-parity multipoles, and clarified how the ASOC is induced in each case. 
The results are useful to predict various unconventional phenomena, such as the spin and valley splitting in the band structure, asymmetric band modulation with a band bottom shift, and peculiar off-diagonal responses including spin and valley Hall effects and ME effects~\cite{hayami2016emergent}. 
The predictions will also be useful to identify the type of staggered ordering by measurement of such phenomena. 
Although the theory provides an archetype of the ASOC physics, it is desired to apply it to realistic situations and to test the predictive power. 

In the present study, we theoretically predict the emergence of the odd-parity multipoles, the ASOC, and the linear ME effects on a centrosymmetric diamond structure, with $5d$ transition metal oxides $A$OsO$_4$ ($A=$ K, Rb, and Cs) in mind~\cite{comment_Yamaura,Song_PhysRevB.90.245117}. 
We derive an effective model for the relevant $e_g$ orbitals by incorporating orbital-dependent hoppings, crystalline electric fields, and the effective SOC. 
The effective SOC in the $e_{g}$ orbitals is obtained by taking into account the atomic SOC with $t_{2g}$ orbitals, and plays an important role in $A$OsO$_4$ as the cubic crystalline electric field is comparable to the atomic SOC~\cite{Song_PhysRevB.90.245117}. 
We classify the staggered spin and orbital orders and the associated odd-parity multipoles from the symmetry point of view. 
For each case, we elucidate what type of ASOC is induced and 
what type of ME effect occurs. 
The results are discussed for $A$OsO$_4$, in which the order parameters are not yet fully identified~\cite{comment_Yamaura,Song_PhysRevB.90.245117}. 
We show that the ME responses are sensitive to the directions of ordered magnetic moments.
This would be useful for KOsO$_4$ and RbOsO$_4$ where an antiferromagnetic order is anticipated but the direction of the moment is unknown. 
We also discuss the results for other electronic orderings by considering orbital and spin-orbital channels, which provides a reference to identify the unknown order parameter in CsOsO$_4$. 

The rest of the paper is organized as follows. 
In Sec.~\ref{sec:Two-orbital model and odd-parity multipoles}, after constructing the effective $e_g$-orbital Hamiltonian, we present possible odd-parity multipoles induced by staggered electronic orderings. 
In Sec.~\ref{sec:Antisymmetric spin-orbit coupling in $e_g$ orbitals}, we show what type of the ASOC is activated in each staggered ordered state. 
In Sec.~\ref{sec: Magnetoelectric Effect}, we elaborate the linear ME effects induced by the odd-parity multipoles associated with staggered orderings. 
Section~\ref{sec:Summary} is devoted to the summary and future perspectives. 

\section{Two-orbital model and odd-parity multipoles}
\label{sec:Two-orbital model and odd-parity multipoles}
In this section, we present an effective model which includes the ASOC hidden on the diamond structure. 
In Sec.~\ref{sec:Hamiltonian}, we derive the effective Hamiltonian for twofold $e_g$ orbitals modulated by the strong SOC through the mixing with $t_{2g}$ orbitals. 
After classifying possible order parameters by symmetry in Sec.~\ref{sec:Odd-parity multipoles due to staggered orderings}, we discuss emergent odd-parity multipoles associated with the staggered electronic orders in Sec.~\ref{sec:Odd-parity multipoles}. 

\subsection{Effective two-orbital Hamiltonian}
\label{sec:Hamiltonian}

\begin{figure}[hbt!]
\begin{center}
\includegraphics[width=1.0 \hsize]{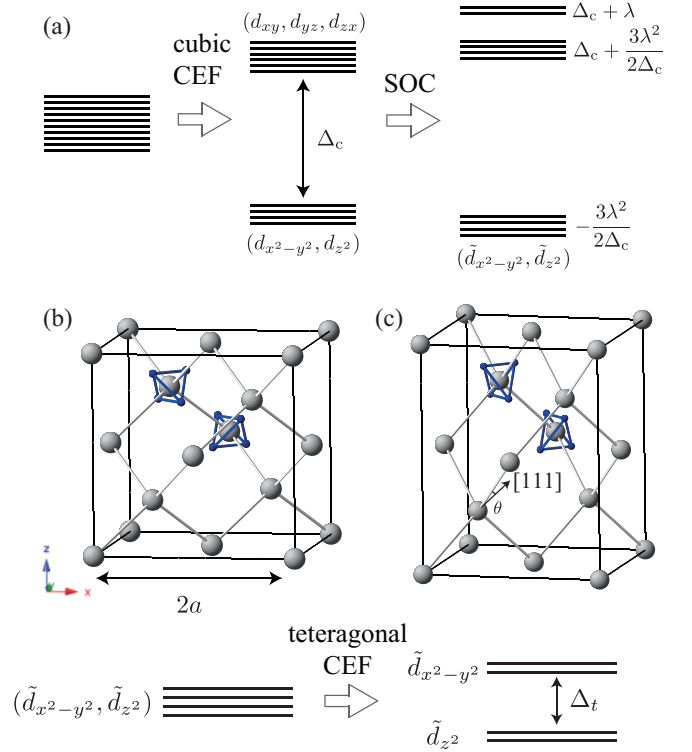} 
\caption{
\label{Fig:diamond_structure}
(a) Schematic picture of the atomic energy levels for 5$d$ orbitals. 
The cubic CEF and SOC represent the cubic crystalline electric field $\Delta_c$ and the atomic spin-orbit coupling $\lambda$, respectively. 
We take into account the lowest-energy levels $\tilde{d}_{x^2-y^2}$ and $\tilde{d}_{z^2}$ [Eqs.~(\ref{eq:basis1}) and (\ref{eq:basis2})], which are derived from the $e_g$ orbitals $d_{x^2-y^2}$ and  $d_{z^2}$ with a modulation through the mixing between $e_g$ and $t_{2g}$ orbitals by the SOC. 
(b), (c) Schematic pictures of the diamond structure in the (b) absence and (c) presence of a tetragonal distortion. 
The lowest panel shows the schematic energy levels for the $\tilde{d}_{x^2-y^2}$ and $\tilde{d}_{z^2}$ orbitals in the absence and presence of the tetragonal CEF $\Delta_t$. 
In (b) and (c), schematic pictures of OsO$_4$ tetrahedra are partly shown. 
}
\end{center}
\end{figure}

Let us consider the energy levels of $d$ electrons on the diamond structure [Fig.~\ref{Fig:diamond_structure}(b)], with 5$d$ transition metal oxides $A$OsO$_4$ ($A=$ K, Rb, and Cs) in mind. 
Under the cubic crystalline electric field $\Delta_c$ in each OsO$_4$ tetrahedron [see Fig.~\ref{Fig:diamond_structure}(b)], the tenfold atomic energy levels are split into the sixfold $t_{2g}$ orbitals and the fourfold $e_g$ orbitals, as shown in the leftmost and middle panels of Fig.~\ref{Fig:diamond_structure}(a). 
Considering the $d^1$ electron configuration in each Os$^{7+}$ cation in $A$OsO$_4$, we focus on the lower-energy $e_g$ orbitals in the following.  

Although the $e_g$ orbitals have no matrix elements for the atomic SOC, they are modulated by taking into account the atomic SOC with the $t_{2g}$ orbitals. 
This becomes relevant when the ratio $\lambda/\Delta_c$ is not negligibly small ($\lambda$ is the coupling constant of the atomic SOC). In fact, a relatively large value of $\lambda/\Delta_c \sim 0.18$ was reported for KOsO$_4$ from the first-principles calculation~\cite{Song_PhysRevB.90.245117}. 
When we take into account such a SOC effect in the second-order perturbation, the $e_g$ and $t_{2g}$ levels are further split: the $e_g$ levels are lowered while keeping the fourfold degeneracy, whereas the $t_{2g}$ levels are split into twofold and fourfold [see the rightmost panel in Fig.~\ref{Fig:diamond_structure}(a)]. 
The $e_g$ orbital bases are modulated by the mixing with the split $t_{2g}$ orbitals as
\begin{align}
\label{eq:basis1}
|\tilde{d}_{x^2 - y^2 \sigma} \rangle &= N \left[ 
| d_{x^2-y^2 \sigma} \rangle 
\mp 2 i \Lambda | d_{xy \sigma} \rangle \right. \nonumber\\
&\qquad\qquad\qquad\quad \left. + i \Lambda | d_{yz \bar{\sigma}} \rangle
\mp\Lambda | d_{zx  \bar{\sigma}} \rangle 
\right],
\\
\label{eq:basis2}
|\tilde{d}_{z^2 \sigma} \rangle &= N \left[ 
| d_{z^2 \sigma} \rangle 
+ \sqrt{3}i \Lambda | d_{yz \bar{\sigma}} \rangle
\pm \sqrt{3}\Lambda | d_{zx\bar{\sigma}} \rangle
\right],
\end{align}
where $|d_{xy\sigma}\rangle$, $|d_{yz\sigma}\rangle$, and $|d_{zx\sigma}\rangle$ ($|d_{x^2-y^2\sigma}\rangle$ and $|d_{z^2\sigma}\rangle$) are the bases for the $t_{2g}$ ($e_g$) orbitals with spin $\sigma=\uparrow$ or $\downarrow$ in the absence of SOC, and $\Lambda$ describes their mixing: 
\begin{equation}
\Lambda = \lambda/(2\Delta_{c}) + \lambda^2/(2\Delta_{c})^2+ \mathcal{O}(\lambda^3/\Delta^3_{c}),
\end{equation} 
for $\lambda < \Delta_c$. 
In Eqs.~(\ref{eq:basis1}) and (\ref{eq:basis2}), 
$N=1/\sqrt{1+6\Lambda^2}$ is the normalization factor and 
$\bar{\sigma}$ represents the opposite spin to $\sigma$. 
The axis of $d$ orbitals is taken along with the crystal axis shown in Fig.~\ref{Fig:diamond_structure}(b). 

We construct a tight-binding model for the bases in Eqs.~(\ref{eq:basis1}) and (\ref{eq:basis2}) by adopting the Slater-Koster parameters for the hopping elements~\cite{Slater_PhysRev.94.1498}. 
The Hamiltonian is given by  
\begin{align}
\label{eq:Ham0cub}
\mathcal{H}^{c}_0 = &-t_0 \sum_{\mathbf{k} \alpha \sigma \sigma'} \left( \gamma_{0\mathbf{k}} c_{{\rm A} \mathbf{k} \alpha \sigma}^{\dagger} c_{{\rm B} \mathbf{k} \alpha \sigma'} + {\rm H.c.}\right) \nonumber \\
&-t_1 \sum_{\substack{\mathbf{k} \alpha\beta \sigma \sigma' \\ \mu=x,y,z}} \left( \gamma_{\mu\mathbf{k}} \left[\tau_{y}\right]_{\alpha\beta} \left[\sigma_{\mu}\right]_{\sigma\sigma'} c_{{\rm A} \mathbf{k} \alpha \sigma}^{\dagger} c_{{\rm B} \mathbf{k} \beta \sigma'} + {\rm H.c.}\right),  
\end{align} 
where $c_{s \mathbf{k} \alpha \sigma}^{\dagger}$ ($c_{s \mathbf{k} \alpha \sigma}$) is the creation (annihilation) operator of a conduction electron with sublattice $s=$ A or B, orbital $\alpha$ ($\alpha=1$ and $2$ correspond to $\tilde{d}_{x^2-y^2}$ and $\tilde{d}_{z^2}$, respectively), and spin $\sigma$ at wave number $\mathbf{k}$. 
The first and second terms in Eq.~(\ref{eq:Ham0cub}) represent the kinetic energy of the conduction electrons from the intra- and inter-orbital hoppings, respectively. 
In the second term, the Pauli matrices $\bm{\sigma} = (\sigma_x, \sigma_y, \sigma_z)$ and $\bm{\tau} = (\tau_x, \tau_y, \tau_z)$ are introduced to describe the spin and ``orbital'' degrees of freedom, respectively.
Note that this orbital degree of freedom in the $e_g$ orbital bases represents higher-order electric and magnetic multipoles: $\tau_{x}$ and $\tau_{z}$ correspond to the electric quadrupoles, $x^{2}-y^{2}$ and $3z^{2}-r^{2}$, respectively, while $\tau_{y}$ represents the magnetic octupole, $l_{x}l_{y}l_{z}$. 
The spin and orbital dependences in the inter-orbital hoppings originate from the mixing between the original $e_g$ and $t_{2g}$ orbitals by the effective SOC. 
Indeed, $t_1$ is proportional to $\Lambda$ and vanishes in the absence of the SOC. 
In the following, we take the lattice constant $a=1$ as the length unit [see Fig.~\ref{Fig:diamond_structure}(b)], and restrict the sums in Eq.~(\ref{eq:Ham0cub}) to the nearest-neighbor sites on the diamond structure. 
Then, the wave number dependences of the hoppings between the same orbitals $t_0$ and the different orbitals $t_1$ are given by 
\begin{align}
\gamma_{0\mathbf{k}} & = 4 \left(\cos \frac{k_x}{2} \cos \frac{k_y}{2}\cos \frac{k_z}{2} + i \sin \frac{k_x}{2}\sin \frac{k_y}{2}\sin \frac{k_z}{2}\right), \\
\gamma_{x \mathbf{k}} & = -4 \left(\cos \frac{k_x}{2} \sin \frac{k_y}{2}\sin \frac{k_z}{2} + i \sin \frac{k_x}{2}\cos \frac{k_y}{2}\cos \frac{k_z}{2}\right), \\
\gamma_{y \mathbf{k}} & = -4 \left(\sin \frac{k_x}{2} \cos \frac{k_y}{2}\sin \frac{k_z}{2} + i \cos \frac{k_x}{2}\sin \frac{k_y}{2}\cos \frac{k_z}{2}\right),  \\
\gamma_{z \mathbf{k}} & = -4 \left(\sin \frac{k_x}{2} \sin \frac{k_y}{2}\cos \frac{k_z}{2} + i \cos \frac{k_x}{2}\cos \frac{k_y}{2}\sin \frac{k_z}{2}\right). 
\end{align}
The hoppings consist of the symmetric and antisymmetric parts with respect to $\mathbf{k}$, and the latter is essential to generate the ASOC. 

In addition, we also consider the effect of a tetragonal distortion [see Fig.~\ref{Fig:diamond_structure}(c)], which is indeed observed in $A$OsO$_4$~\cite{comment_Yamaura}. 
Under the tetragonal crystalline electric field, the energy levels for $\tilde{d}_{x^2-y^2}$ and $\tilde{d}_{z^2}$ orbitals are split by $\Delta_t$, as shown in the lower panel of Figs.~\ref{Fig:diamond_structure}(b) and \ref{Fig:diamond_structure}(c). 
At the same time, the hopping integrals are modulated by the lattice distortion. 
Then, we describe the effect of the tetragonal distortion by 
the additional terms to $\mathcal{H}^c_0$ in Eq.~(\ref{eq:Ham0cub}), which are given by 
\begin{align}
\label{eq:Ham0tet}
\mathcal{H}^{t}_0 = &-\delta t_0 \sum_{\mathbf{k} \alpha  \beta \sigma \sigma'} \left( \gamma_{0\mathbf{k}} \left[\tau_{z}\right]_{\alpha\beta} c_{{\rm A} \mathbf{k} \alpha \sigma}^{\dagger} c_{{\rm B} \mathbf{k} \alpha \sigma'} + {\rm H.c.}\right) \nonumber \\
&+\frac{\delta t_1}{2} \sum_{\substack{\mathbf{k}\alpha\beta \sigma \sigma'\\ \mu=x,y}} \left( \gamma_{\mu\mathbf{k}} \left[\tau_{y}\right]_{\alpha\beta}\left[\sigma_{\mu}\right]_{\sigma\sigma'} c_{{\rm A} \mathbf{k} \alpha \sigma}^{\dagger} c_{{\rm B} \mathbf{k} \beta \sigma'} + {\rm H.c.}\right) \nonumber \\
&-\delta t_1 \sum_{\mathbf{k} \alpha\beta \sigma \sigma'} \left( \gamma_{z\mathbf{k}} \left[\tau_{y}\right]_{\alpha\beta} \left[\sigma_{z}\right]_{\sigma\sigma'} c_{{\rm A} \mathbf{k} \alpha \sigma}^{\dagger} c_{{\rm B} \mathbf{k} \beta \sigma'} + {\rm H.c.}\right) \nonumber \\
&+\Delta_t \sum_{i} \sum_{\alpha \beta}  \sum_{\sigma} \left[\tau_z\right]_{\alpha \beta} c^{\dagger}_{i \alpha \sigma} c_{i \beta \sigma}.   
\end{align} 
In the first three terms, we take into account the effect of the tetragonal distortion on hopping elements by considering a slight change of the bond direction from the $\langle 111 \rangle$ direction by a small angle $\theta$, as shown in Fig.~\ref{Fig:diamond_structure}(c): 
we take into account the lowest order of $\theta$, namely, $\delta t_0, \delta t_1 \propto \theta$.
The fourth term in Eq.~(\ref{eq:Ham0tet}) represents the tetragonal crystal field splitting between the $\tilde{d}_{x^2-y^2}$ and $\tilde{d}_{z^2}$ levels. 

\subsection{Staggered electronic orders} 
\label{sec:Odd-parity multipoles due to staggered orderings}

In the present study, we examine the electronic states of the system in the presence of spontaneous symmetry breaking by staggered electronic orders on the bipartite diamond structure.
Considering the experiments for $A$OsO$_4$ in which the magnetic susceptibility shows an anomaly~\cite{comment_Yamaura}, we focus on all possible states with breaking of time-reversal symmetry. 
Such order parameters are represented by the Pauli matrices for spin ($\bm{\sigma}$) and orbital ($\bm{\tau}$) indices: three spin orders $\sigma_\mu$ ($\mu=x,y,z$), an ``orbital'' order $\tau_y$ (magnetic octupole), and six spin-orbital orders $\sigma_\mu \tau_\nu$ ($\mu=x,y,z$ and $\nu=x,z$). 
Moreover, we focus on the simplest realizations of the breaking of spatial inversion symmetry (parity breaking) by adopting the staggered-type orders on the bipartite structure, as these ten symmetry-broken states exhibit even parity with respect to spatial inversion.
We introduce a mean-field term,
\begin{align}
\label{eq:Ham1}
\mathcal{H}_1 = -h \sum_{s \mathbf{k} \alpha \beta \sigma \sigma'} c^{\dagger}_{s \mathbf{k} \alpha \sigma} p(s) \left[\sigma_{\tilde{\mu}}\right]_{\sigma \sigma'} \left[\tau_{\tilde{\nu}}\right]_{\alpha \beta}
c_{s \mathbf{k}\beta\sigma'}, 
\end{align}
where $\tilde{\mu}, \tilde{\nu}=0,x,y,z$, and $\sigma_0$ and $\tau_0$ are the unit matrices in spin and orbital spaces. 
$h$ is the magnitude of the symmetry-breaking field and $p(s)=+1(-1)$ for $s=$ A (B). 
Thus, the total mean-field Hamiltonian is given by $\mathcal{H} = \mathcal{H}^c_0 + \mathcal{H}^t_0 + \mathcal{H}_1$. 

Each order parameter can be locally represented by taking a product of the modulated $e_g$ bases in Eqs.~(\ref{eq:basis1}) and (\ref{eq:basis2}) supplemented by the spin $1/2$ basis. 
When the system preserves the cubic symmetry in the absence of the tetragonal distortion ($\mathcal{H}^t_0=0$), the product of the $e_g$ orbital bases with spin is reduced under the point group $T_d$ as 
\begin{multline}
\label{eq:product_cub}
(E \otimes D_{1/2}) \otimes (E \otimes D_{1/2}) \\
= A_1^{+} \oplus A_2^{-} \oplus E^{+} \oplus 2 T_1^{-} \oplus T_2^{+}\oplus T_2^{-}, 
\end{multline}
where $E$ and $D_{1/2}$ in the first line represent the $e_g$ orbital and spin $1/2$, respectively, and the superscripts $+$ and $-$ in the second line represent time-reversal property. 
Note that the irreducible representations with the superscript $-$ correspond to the order parameters with breaking of time-reversal symmetry while keeping even parity.
These ten order parameters listed above are classified by the irreducible representations in Eq.~(\ref{eq:product_cub}) as follows: 
$\tau_y$ belongs to $A^-_2$, 
$\sigma_\mu$ and $\sigma_\mu \tau_z$ to $T^-_1$, and 
$\sigma_\mu \tau_x$ to $T^-_2$, respectively ($\mu=x,y,z$).  
The order parameters are accompanied with the odd-parity multipoles by aligning in a staggered way, as described in Sec.~\ref{sec:Odd-parity multipoles}.
As we restrict ourselves to the magnetic order parameters, we will omit the superscript $-$ in what follows. 
The result is summarized in Table~\ref{tab:cubic}. 

\begin{table}[htb!]
\begin{center}
\caption{
Classification of the time-reversal breaking even-parity order parameters (OPs) and associated odd-parity multipoles under the cubic $T_d$ and tetragonal $D_{2d}$ symmetries. 
Some of the odd-parity multipoles in the momentum space, defined as Eq.~(\ref{eq:fT1})-(\ref{eq:fT3}) and (\ref{eq:fT1_prime})-(\ref{eq:fT3_prime}), are also shown. 
}
\label{tab:cubic}
\begingroup
\renewcommand{\arraystretch}{1.2}
 \begin{tabular}{cccccc|c|c|c|c|c|c|c|c|c|c|c|c|c|c|c|c|}
 \hline\hline
$T_d$ & \ \ $D_{2d}$ \ \ & \multicolumn{2}{c}{
even-parity OP}  & odd-parity multipole  \\
\hline
$A_{1}$ &\ \ $A_{1}$ \ \ & \multicolumn{2}{c}{---}   & --- \\ \hline
$A_{2}$ &\ \ $B_{1}$ \ \ & \multicolumn{2}{c}{$\tau_y$} &$M_{0}$ \\ \hline
\multirow{2}{*}{$E$} &\ \ $A_{1}$\ \ & \multicolumn{2}{c}{---}  & $M_{v}$ \\ 
& \ \ $B_{1}$ \ \ & \multicolumn{2}{c}{---} & $M_{u}$
\\ \hline
\multirow{2}{*}{$T_{1}$} &
\ \ $A_{2}$ \ \ & $\sigma_z$ & $\sigma_z \tau_z$ & $M_{xy}$\ \ \ \  \ \ \  \ \ \  \ \ \  \ \ \ $f_{z}(\mathbf{k})$
 \\ 
&
\ \ $E$ \ \ & $(\sigma_x,\sigma_y)$ & $(\sigma_x \tau_z,\sigma_y \tau_z)$ & $(M_{yz},M_{zx})$\ \ \  $(f_{x}(\mathbf{k}),f_{y}(\mathbf{k}))$
\\
\hline
\multirow{2}{*}{$T_{2}$} &
\ \ $B_{2}$ \ \ &\multicolumn{2}{c}{$\sigma_z \tau_x$}  &$T_z$\ \ \ \  \ \ \  \ \ \  \ \ \  \ \ \ $f_{z}'(\mathbf{k})$
\\ 
&
\ \ $E$ \ \ &  \multicolumn{2}{c}{$(\sigma_x \tau_x,\sigma_y \tau_x)$}  & $(T_x,T_y)$\  \ \ ($f_{x}'(\mathbf{k}),f_{y}'(\mathbf{k})$)
\\ \hline\hline
\end{tabular}
\endgroup
\end{center}
\end{table}

Meanwhile, in the presence of a tetragonal distortion, the symmetry is lowered from $T_d$ to $D_{2d}$. 
Using the reduction rule, $A_{2}\to B_{1}$, $E\to (A_{1},B_{1})$, $T_{1}\to (A_{2},E)$, and $T_{2}\to (B_{2},E)$, the ten order parameters are classified by the irreducible representations of the $D_{2d}$ point group as follows: 
$\sigma_z$ and $\sigma_z \tau_z$ belong to $A_2$, 
$\tau_y$ belongs to $B_1$, 
$\sigma_z \tau_x$ to $B_2$, and 
($\sigma_x$,$\sigma_y$), ($\sigma_x \tau_z$,$\sigma_y \tau_z$), and ($\sigma_x \tau_x$,$\sigma_y \tau_x$) to $E$. 
The result is also summarized in Table~\ref{tab:cubic}. 

\subsection{Odd-parity multipoles}
\label{sec:Odd-parity multipoles}

The staggered electronic orders simultaneously induce odd-parity multipoles, as they break spatial inversion symmetry in addition to time-reversal symmetry. 
Two types of odd-parity multipoles can be defined under the time-reversal symmetry breaking in the expansion of the electromagnetic vector potential~\cite{dubovik1986axial,Kusunose_JPSJ.77.064710,kuramoto2009multipole,Spaldin_0953-8984-20-43-434203}. 
One is magnetic multipoles, which are defined as  
\begin{align}
\label{eq:mag_multipole}
M_{lm}&= - \mu_{\rm B} \sum_j\left(\frac{2 \mathbf{l}_j}{l+1}+\bm{\sigma}_{j} \right)\cdot \bm{\nabla} O_{lm}(\mathbf{r}_{j}), 
\end{align}
where $\mu_{\rm B}$ is the Bohr magneton, and $\mathbf{l}_j$ and $\bm{\sigma}_j$ stand for orbital and spin angular momenta for an electron at the position $\mathbf{r}_j$, respectively. 
We introduced $O_{lm}(\mathbf{r})=\sqrt{4\pi/(2l+1)}r^{l}Y_{lm}(\mathbf{r})$ with $Y_{lm}$ being the spherical harmonics, and $l$ and $m$ are the azimuthal and magnetic quantum numbers, respectively. 
The other is magnetic toroidal multipoles, which are given by~\cite{comment_hayami} 
\begin{align}
T_{lm}&= -\mu_{\rm B}\sum_j \left[ \frac{\mathbf{r}_j}{l+1} \times \left(\frac{2 \mathbf{l}_j}{l+2}+\bm{\sigma_{j}} \right) \right]\cdot \bm{\nabla} O_{lm}(\mathbf{r}_j).  
\end{align}
As $Y_{lm}$ has parity $(-1)^l$, the lowest-rank magnetic multipoles with parity odd are magnetic quadrupoles: 
\begin{align}
\label{eq:Myz}
M_{yz}& \propto y l_z + z l_y + {\rm H.c.}, \\
\label{eq:Mzx}
M_{zx}& \propto z l_x + x l_z + {\rm H.c.}, \\
\label{eq:Mxy}
M_{xy} &\propto x l_y + y l_x +  {\rm H.c.} , \\
\label{eq:Mz2}
M_{u}& \propto 3 z l_z  -x l_x - y l_y + {\rm H.c.},  \\
\label{eq:Mx2}
M_{v}& \propto x l_x - y l_y + {\rm H.c.},
\end{align}
and toroidal dipoles: 
\begin{align}
\label{eq:toroidal}
\mathbf{T}\propto \mathbf{r} \times \mathbf{l} + {\rm H.c.}, 
\end{align}
where we omit the site index and spin angular momenta for simplicity. 
We also define a magnetic monopole $M_{0}$ as a pseudoscalar, which is given by 
\begin{align}
\label{eq:Magmono}
M_{0} \propto x l_x  + y l_y + z l_z + {\rm H.c.}. 
\end{align}
Note that the magnetic monopole does not appear in the expansion of the magnetic vector potential in Eq.~(\ref{eq:mag_multipole}). Nevertheless, it may be activated when the the magnetic unit cell possesses the same symmetry as the pseudoscalar. 

In the momentum space, these odd-parity multipoles are given by~\cite{Watanabe_PhysRevB.96.064432}
\begin{align}
\label{eq:fT1}
M_{yz}& \propto (\cos k_{y}-\cos k_{z})\sin k_{x}=f_{x}(\mathbf{k}),\\
\label{eq:fT2}
M_{zx}& \propto (\cos k_{z}-\cos k_{x})\sin k_{y}=f_{y}(\mathbf{k}), \\
\label{eq:fT3}
M_{xy} &\propto (\cos k_{x}-\cos k_{y})\sin k_{z}=f_{z}(\mathbf{k}), \\
\label{eq:Mz2k}
M_{u}& \propto \sin k_{x} \sin k_{y} \sin k_{z}(\cos k_{x}-\cos k_{y}), \\
\label{eq:Mx2k}
M_{v}& \propto \sin k_{x} \sin k_{y} \sin k_{z},
\end{align}
and
\begin{align}
\label{eq:fT1_prime}
T_{x}\propto (\cos k_{y}+\cos k_{z})\sin k_{x}=f_{x}'(\mathbf{k}), \\ 
\label{eq:fT2_prime}
T_{y}\propto (\cos k_{z}+\cos k_{x})\sin k_{y}=f_{y}'(\mathbf{k}), \\ 
\label{eq:fT3_prime}
T_{z}\propto (\cos k_{x}+\cos k_{y})\sin k_{z}=f_{z}'(\mathbf{k}).
\end{align}
Note that they are time-reversal and spatial inversion odd.
The asymptotic forms of Eqs.~(\ref{eq:fT1})-(\ref{eq:fT3}) in the $\mathbf{k} \to \mathbf{0}$ limit are given by 
\begin{align}
\label{eq:fT1limit}
f_{x} (\mathbf{k}) &\to (k_y^2 - k_z^2) k_x, \\
\label{eq:fT2limit}
f_{y} (\mathbf{k}) &\to ( k_z^2 - k_x^2 )  k_y, \\
\label{eq:fT3limit}
f_{z} (\mathbf{k}) &\to ( k_x^2 - k_y^2 )  k_z.  
\end{align}
We note that they share the functional forms with the Dresselhaus-type ASOC appearing in the diamond structure; see Sec.~\ref{sec:Antisymmetric spin-orbit coupling in paramagnetic state}.
Similarly, the asymptotic forms of Eqs.~(\ref{eq:fT1_prime})-(\ref{eq:fT3_prime}) in the $\mathbf{k} \to \mathbf{0}$ limit are given by 
\begin{align}
\label{eq:fT1limit_1st}
f_{x}' (\mathbf{k}) &\to k_x, \\
\label{eq:fT2limit_1st}
f_{y}' (\mathbf{k}) &\to  k_y, \\
\label{eq:fT3limit_1st}
f_{z}' (\mathbf{k}) &\to  k_z,   
\end{align}
which give $\mathbf{k}$-linear contributions.

These odd-parity multipoles are induced by the staggered even-parity order parameters within the same irreducible representation. 
Under the point group $T_d$ in the cubic symmetry, the order parameter $\tau_y$ activates the magnetic monopole $M_{0}$. 
Meanwhile, $\sigma_\mu \tau_\nu$ ($\mu=x,y,z$ and $\nu=0, z$) activates the magnetic quadrupoles $M_{yz}$, $M_{zx}$, and $M_{xy}$; $\sigma_\mu \tau_x$ ($\mu=x,y,z$) activate the toroidal dipoles $T_x$, $T_y$, and $T_z$. 
The result is summarized in Table~\ref{tab:cubic}. 

On the other hand, under the point group $D_{2d}$ in the tetragonal symmetry, 
$\tau_y$ activates a linear combination of $M_{0}$ and $M_{u}$, since $B_{1}$ of the point group $D_{2d}$ is reduced from $A_{2}$ and $E$ of the point ground $T_d$.
Meanwhile, $\sigma_z$ and $\sigma_z \tau_z$ activate the magnetic quadrupole $M_{xy}$. 
$\sigma_z \tau_x$ activates the toroidal dipole $T_z$, while $\sigma_\mu \tau_\nu$ 
($\mu=x,y$ and $\nu=0,x,z$) activate a linear combination of $(M_{yz}, M_{zx})$ and $(T_x, T_y)$, as $E$ of the point group $D_{2d}$ is reduced from $T_{1}$ and $T_{2}$ of the point group of $T_d$.
The result is also summarized in Table~\ref{tab:cubic}.

\section{Antisymmetric spin-orbit coupling in $e_g$ orbitals}
\label{sec:Antisymmetric spin-orbit coupling in $e_g$ orbitals}
In this section, we discuss the ASOC in our $e_g$-orbital model arising from the effective SOC by the $e_g$-$t_{2g}$ mixing. 
In Sec.~\ref{sec:Antisymmetric spin-orbit coupling in paramagnetic state}, we show that the ASOC is hidden in a staggered form in the paramagnetic state. 
Then, in Sec.~\ref{sec:Antisymmetric spin-orbit coupling in ordered states}, we show a net component of the ASOC is induced by the staggered electronic orderings discussed in the previous section. 
We discuss how the form of ASOC depends on the electronic order.

In order to obtain the explicit form of the ASOC, we perform the canonical transformation at one of two sublattices by tracing out the other sublattice degree of freedom. 
In other words, we treat the electron transfers between different sublattices in $\mathcal{H}^c_0+\mathcal{H}^t_0$ as the perturbation. 
The details of the scheme are described in Ref.~\onlinecite{hayami2016emergent}. 

\subsection{Paramagnetic state}
\label{sec:Antisymmetric spin-orbit coupling in paramagnetic state}

In the paramagnetic state, there is no net component of the ASOC, but a staggered component is allowed due to the absence of inversion symmetry at each lattice site. 
In other words, the effective ASOC has the same amplitude but an opposite sign between the two sublattices. 
Indeed, in the absence of a tetragonal distortion, the ASOC is obtained as 
\begin{align}
\label{eq:paraASOC1cub}
\mathcal{H}^{\rm para}_{\rm ASOC} (\mathbf{k}) &\propto  4 t^2_1 \left[  f_{x} (\mathbf{k})    \sigma_x +   f_{y} (\mathbf{k})   \sigma_y +  f_{z} (\mathbf{k})  \sigma_z  \right] \rho_z \tau_0, 
\end{align}
where $\rho_z$ is the $z$ component of the Pauli matrix, representing that the ASOC is staggered between the two sublattices. 
In Eq.~(\ref{eq:paraASOC1cub}), $f_\mu(\mathbf{k})$ ($\mu=x,y,z$) are defined in Eqs.~(\ref{eq:fT1})-(\ref{eq:fT3}), and hence, the asymptotic form of the effective ASOC in the $\mathbf{k} \to \mathbf{0}$ limit is similar to the Dresselhaus-type ASOC, preserving the threefold rotational symmetry around the $\langle 111 \rangle$ direction~\cite{Dresselhaus_PhysRev.100.580}, as expected for the diamond structure. 

Meanwhile, in the presence of a tetragonal distortion, the ASOC in Eq.~(\ref{eq:paraASOC1cub}) is deformed as 
\begin{align}
\label{eq:paraASOC1tetra}
&\mathcal{H}^{\rm para (1)}_{\rm ASOC} (\mathbf{k}) \nonumber \\
&\propto   \tilde{t}_a \left[   \tilde{t}_b \left\{f_{x} (\mathbf{k})    \sigma_x +    f_{y} (\mathbf{k})   \sigma_y \right\} + 
\tilde{t}_a f_{z} (\mathbf{k})  \sigma_z  \right] \rho_z \tau_0, 
\end{align}
where $\tilde{t}_a=2t_1-\delta t_1$ and $\tilde{t}_b= 2( t_1 +  \delta t_1)$. 
In addition to these terms, other components are also induced by the tetragonal distortion: 
\begin{align}
\label{eq:paraASOC2tetra}
&\mathcal{H}^{\rm para (2)}_{\rm ASOC} (\mathbf{k}) \nonumber \\  
&\propto \frac{  t_0    }{\Delta_t} \left[ \tilde{t}_a \left\{ f_{x}' (\mathbf{k})  \sigma_x  
+    f_{y}' (\mathbf{k})  \sigma_y  \right\} 
+   \tilde{t}_b  f_{z}' (\mathbf{k})  \sigma_z \right]\rho_z\tau_x \nonumber \\
&+ \frac{ \tilde{t}_a  }{\Delta_t} \left[\tilde{t}_b \left\{ f_{x} (\mathbf{k})   \sigma_x  
+     f_{y} (\mathbf{k}) \sigma_y \right\}
+  \tilde{t}_a  f_{z} (\mathbf{k}) \sigma_z\right] \rho_z \tau_z, 
\end{align}
where $f_\mu'(\mathbf{k})$ ($\mu=x,y,z$) are defined in Eqs.~(\ref{eq:fT1_prime})-(\ref{eq:fT3_prime}).
These terms appear due to the mixing of $T_{1}$ and $T_{2}$ irreducible representations of $T_{d}$ symmetry under $D_{2d}$ symmetry. 

These staggered ASOCs often lead to anomalous magnetotransport phenomena even in the paramagnetic state, such as the (spin) Hall effect~\cite{Fu_PhysRevLett.98.106803}. 
In fact, an anisotropic anomalous Hall effect was observed under the magnetic field in the spinel compound FeCr$_2$S$_4$~\cite{ohgushi2006anisotropic}, whose origin is understood from the ASOC similar to Eq.~(\ref{eq:paraASOC1cub}). 
It would be intriguing to investigate magnetotransport phenomena under a tetragonal distortion, since our analysis indicates that the effective two-orbital model includes additional contributions, as shown in Eqs.~(\ref{eq:paraASOC1tetra}) and (\ref{eq:paraASOC2tetra}). 
This suggests that further interesting phenomena may arise from not only a spontaneous lattice distortion but also an applied pressure. 
Such investigation is left for future study. 

\subsection{Ordered states}
\label{sec:Antisymmetric spin-orbit coupling in ordered states}

In this section, we turn to the effective ASOC induced by the staggered electronic orderings discussed in Sec.~\ref{sec:Odd-parity multipoles due to staggered orderings}. 
We use a similar procedure to the previous subsection for the model $\mathcal{H}^c_0+\mathcal{H}^t_0+\mathcal{H}_1$, where $\mathcal{H}_1$ in Eq.~(\ref{eq:Ham1}) represents any of the ten order parameters. 
Among many contributions, we focus on the ASOCs proportional to the orbital component $\tau_0$ and $\tau_y$, since they lead to the linear ME effects, as discussed in Sec.~\ref{sec: Magnetoelectric Effect}. 

\subsubsection{Spin order}
\label{sec:Spin order}

In the spin ordered states, $\mathcal{H}_1$ is proportional to $\sigma_\mu \tau_0$ ($\mu=x,y,z$). 
Under these parity breaking orders, a net component of the ASOC is induced in the form of 
\begin{align}
\label{eq:ASOC_spinx}
\mathcal{H}^{\sigma_x}_{\rm ASOC} &\propto 
\frac{ h t_0}{\Delta_t^2}
\left[  \tilde{t}_b f_{z}'(\mathbf{k})  \sigma_y
- \tilde{t}_a  f_{y}'(\mathbf{k})  \sigma_z 
\right]\rho_0 \tau_y  
\nonumber  \\
  &+ \frac{ h \tilde{t}_a \tilde{t}_b }{2\Delta_{t}^2} f_{x}(\mathbf{k}) \rho_0 \sigma_0 \tau_0, \\
  \label{eq:ASOC_spiny}
\mathcal{H}^{\sigma_y}_{\rm ASOC} &\propto 
\frac{ h t_0}{\Delta_t^2}
\left[  \tilde{t}_a f_{x}'(\mathbf{k})  \sigma_z
- \tilde{t}_b  f_{z}'(\mathbf{k})  \sigma_x 
\right]\rho_0 \tau_y 
\nonumber  \\
  &+ \frac{ h \tilde{t}_a \tilde{t}_b }{2\Delta_{t}^2} f_{y}(\mathbf{k}) \rho_0 \sigma_0 \tau_0, \\
  \label{eq:ASOC_spinz}
\mathcal{H}^{\sigma_z}_{\rm ASOC} &\propto 
\frac{ h t_0 \tilde{t}_a}{\Delta_t^2}
 \left[   f_{y}'(\mathbf{k})  \sigma_x
-   f_{x}'(\mathbf{k})  \sigma_y 
\right]\rho_0 \tau_y \nonumber \\
&+\frac{ h \tilde{t}_a^2 }{2\Delta_{t}^2}  f_{z}(\mathbf{k}) \rho_0 \sigma_0 \tau_0. 
\end{align}
In contrast to the paramagnetic case, the ASOCs are proportional to the unit matrix in sublattice space $\rho_0$, namely, spatially uniform. 
Moreover, the uniform ASOCs are proportional to $h$, which indicates that they are induced by the spontaneous electronic orderings~\cite{hayami2016emergent}. 
In the presence of the cubic symmetry, the coefficients of Eqs.~(\ref{eq:ASOC_spinx})-(\ref{eq:ASOC_spinz}) are equivalent, while they take different values for $(\sigma_x, \sigma_y)$ and $\sigma_z$ under a tetragonal distortion. 

From the symmetry point of view, the effective ASOCs in Eqs.~(\ref{eq:ASOC_spinx})-(\ref{eq:ASOC_spinz}) break both spatial inversion and time-reversal symmetries, as $f_{\alpha}'(\mathbf{k})  \sigma_\mu \tau_y$ and $f_{\alpha}(\mathbf{k}) \sigma_0 \tau_0$ are odd under both time-reversal and spatial inversion operations. 
This indicates that the effective ASOCs induced by the staggered spin orders lead to the asymmetric band structure~\cite{hayami2016emergent}. 
Furthermore, these ASOCs represent the entanglement between the orbital motion and spin moments, as they include the product of $\mathbf{k}$ and $\bm{\sigma}$. In the spin ordered states, their directions in the first term in Eqs.~(\ref{eq:ASOC_spinx})-(\ref{eq:ASOC_spinz}) are perpendicular to each other; for example, in the $\sigma_x$ ordered state, $k_z$ ($k_y$) is coupled with $\sigma_y$ ($\sigma_z$). 
This indicates the possibility of transverse ME effects, as discussed in Sec.~\ref{sec: Magnetoelectric Effect}. 

\subsubsection{Octupole order}
\label{sec:Orbital order}
In the magnetic octupole ordered state where $\mathcal{H}_1$ is proportional to $ \sigma_0\tau_y$, a uniform ASOC consists of three terms, which is given by 
\begin{align}
\label{eq:ASOC_tauy}
\mathcal{H}^{\tau_y}_{\rm ASOC} &\propto \frac{h \tilde{t}_a}{\Delta^2_t} 
\left[ \tilde{t}_b \left\{ f_{x}(\mathbf{k})  \sigma_x +  f_{y}(\mathbf{k})  \sigma_y \right\} \right. \nonumber\\
&\qquad\qquad\qquad \left. +   \tilde{t}_a f_{z}(\mathbf{k})  \sigma_z 
\right] \rho_0 \tau_y .
\end{align}
Under the cubic symmetry, each term in Eq.~(\ref{eq:ASOC_tauy}) has the same coefficient so that the threefold rotational symmetry around the $\langle 111 \rangle$ axis is preserved, while they take different coefficients for $(\sigma_x, \sigma_y)$ and $\sigma_z$ terms under the tetragonal crystalline electric field. 
In contrast to the ASOCs in the spin ordered states, $\mathbf{k}$ is coupled with $\bm{\sigma}$ in the same component: $k_\mu \sigma_\mu$ ($\mu=x,y,z$). 
This difference leads to different type of the ME responses, i.e., longitudinal ME responses, as discussed in Sec.~\ref{sec: Magnetoelectric Effect}. 

\subsubsection{Spin-orbital order}
\label{sec:Spin-orbital order}
In the spin-orbital ordered states, $\mathcal{H}_1$ is proportional to $ \sigma_\mu\tau_\nu$ ($\mu=x,y,z$ and $\nu=x,z$). 
Among them, the states with the orbital $\tau_z$ component show the following uniform ASOCs:  
\begin{align}
\label{eq:ASOC_spinxtauz}
\mathcal{H}^{\sigma_x \tau_z}_{\rm ASOC} & \propto \frac{h \delta t_0}{\Delta^2_t} \left[\tilde{t}_b f_{z}'(\mathbf{k})  \sigma_y  -   \tilde{t}_a f_{y}'(\mathbf{k})  \sigma_z \right]\rho_0 \tau_y, \\
\label{eq:ASOC_spinytauz}
\mathcal{H}^{\sigma_y \tau_z}_{\rm ASOC} & \propto \frac{h \delta t_0}{\Delta^2_t} \left[  \tilde{t}_a f_{x}'(\mathbf{k})  \sigma_z- \tilde{t}_b f_{z}'(\mathbf{k})  \sigma_x    \right]\rho_0 \tau_y, \\
\label{eq:ASOC_spinztauz}
\mathcal{H}^{\sigma_z \tau_z}_{\rm ASOC} & \propto \frac{h \delta t_0 \tilde{t}_a}{\Delta^2_t} \left[    f_{y}'(\mathbf{k})  \sigma_x  -   f_{x}'(\mathbf{k})  \sigma_y  \right]\rho_0 \tau_y.  
\end{align}
The forms of ASOCs in Eqs.~(\ref{eq:ASOC_spinxtauz})-(\ref{eq:ASOC_spinztauz}) are similar to the first terms in Eqs.~(\ref{eq:ASOC_spinx})-(\ref{eq:ASOC_spinz}), which implies that similar physical properties are obtained between these ordered states. 
The difference is in the coefficients: in the spin-orbital ordered states, the coefficients in Eqs.~(\ref{eq:ASOC_spinxtauz})-(\ref{eq:ASOC_spinztauz}) are proportional to $\delta t_0$, which indicates that the effective ASOCs are induced only in the presence of a tetragonal distortion. 
This is in contrast to the ASOCs in Eqs.~(\ref{eq:ASOC_spinx})-(\ref{eq:ASOC_spinz}), which remain finite even in the cubic system. 

Similarly, the uniform ASOCs in the spin-orbital ordered states with the orbital $\tau_x$ component are represented by 
\begin{align}
\label{eq:ASOC_spinxtaux}
\mathcal{H}^{\sigma_x \tau_x}_{\rm ASOC}  &\propto \frac{ h \delta t_0 \tilde{t}_a }{\Delta_{t}^2}f_{x}'(\mathbf{k}) \rho_0 \sigma_0 \tau_0, \\
\label{eq:ASOC_spinytaux}
\mathcal{H}^{\sigma_y \tau_x}_{\rm ASOC} &\propto \frac{ h \delta t_0 \tilde{t}_a }{\Delta_{t}^2} f_{y}'(\mathbf{k})\rho_0 \sigma_0 \tau_0, \\
\label{eq:ASOC_spinztaux}
\mathcal{H}^{\sigma_z \tau_x}_{\rm ASOC} &\propto \frac{ h \delta t_0 \tilde{t}_b  }{\Delta_{t}^2} f_{z}'(\mathbf{k})\rho_0 \sigma_0 \tau_0.  
\end{align}
There are no spin and orbital components for these ASOCs as in the second terms in Eqs.~(\ref{eq:ASOC_spinx})-(\ref{eq:ASOC_spinz}). 
The ASOCs in Eqs.~(\ref{eq:ASOC_spinxtaux})-(\ref{eq:ASOC_spinztaux}) give rise to a band deformation with the band bottom shift to the $k_\mu$ direction due to the $\mathbf{k}$-linear contribution in Eqs.~(\ref{eq:fT1limit_1st})-(\ref{eq:fT3limit_1st}). 
This is regarded as an effective toroidal field along the $k_\mu$ direction in Eqs.~(\ref{eq:fT1_prime})-(\ref{eq:fT3_prime})~\cite{Hayami_PhysRevB.90.024432,hayami2016emergent}. 
Similar to the spin-orbital ordered states with the $\tau_z$ component, these ASOCs are induced only in the tetragonal system, as the coefficients in Eqs.~(\ref{eq:ASOC_spinxtaux})-(\ref{eq:ASOC_spinztaux}) are proportional to $\delta t_0$.

\section{Magnetoelectric Effect}
\label{sec: Magnetoelectric Effect}

In this section, we discuss the possibility of the linear ME effect under the staggered spin and orbital orderings. As the staggered orders on the diamond structure break both spatial inversion symmetry and time-reversal symmetry, they can give rise to the linear ME effects. 
We present the results for the cubic case in Sec.~\ref{sec:Cubic Symmetry} and for the tetragonal case in Sec.~\ref{sec:Tetragonal Symmetry}. 

Before going into the results, we briefly review the relation between the ME effect and the odd-parity multipoles. 
The linear ME effect is induced by the lowest-order odd-parity multipoles in Eqs.~(\ref{eq:Mxy})-(\ref{eq:Magmono}). This is understood from the expansion of the free energy with respect to the electric field $\mathbf{E}$ and the magnetic field $\mathbf{B}$ up to the second order, which is given by~\cite{LandauLifshitz198001,Spaldin_0953-8984-20-43-434203} 
\begin{align}
\label{eq:freeenergy_EH}
F(\mathbf{E},\mathbf{B}) = F_0 - \frac{\varepsilon_{\mu \nu} E_\mu E_\nu}{8 \pi}
- \frac{\mu_{\mu \nu} B_\mu B_\nu}{8 \pi} - \alpha_{\mu \nu} E_\mu B_\nu. 
\end{align}
Here, $F_0$ is the free energy in the absence of the electric and magnetic fields; $\varepsilon_{\mu\nu}$, $\mu_{\mu \nu}$, and $\alpha_{\mu\nu}$ are the dielectric permittivity, magnetic permeability, and ME tensor, respectively. 
The last term in Eq.~(\ref{eq:freeenergy_EH}) is related to the linear ME responses, which is decomposed into three terms: 
\begin{align}
\label{eq:ME_expand}
&\alpha_{\mu \nu} E_\mu B_\nu \nonumber \\ 
&= -M_{0} (\mathbf{E}\cdot \mathbf{B}) - \mathbf{T}\cdot (\mathbf{E}\times \mathbf{B}) - M_{\mu\nu} (E_\mu B_\nu + E_\nu B_\mu). 
\end{align}
The coefficients of each term in Eq.~(\ref{eq:ME_expand}) represent the pseudoscalar magnetic monopole in Eq.~(\ref{eq:Magmono}), toroidal dipole (polar vector) in Eq.~(\ref{eq:toroidal}), and magnetic quadrupole (symmetric traceless pseudotensor) in Eqs.~(\ref{eq:Myz})-(\ref{eq:Mxy}) and the diagonal components related with $M_u$ and $M_v$ in Eqs.~(\ref{eq:Mz2}) and (\ref{eq:Mx2}) as
\begin{align}
&M_{u} = 2M_{zz}-M_{xx}-M_{yy}, \nonumber \\
&M_{v} = M_{xx}-M_{yy}.
\label{eq:M_Q}
\end{align}
Thus, the linear ME effect takes place when the odd-parity multipoles are present. 
We describe it for each case in the following.

In the presence of the magnetic monopole, an isotropic longitudinal ME responses are activated as 
\begin{align}
\label{eq:ME_monopole}
\mathbf{P} \propto M_{0}  \mathbf{B},  \ \ \ 
\mathbf{M} \propto M_{0}  \mathbf{E}, 
\end{align}
where $\mathbf{P}$ and $\mathbf{M}$ are the induced electric polarization and magnetization, respectively. Thus, the electric polarization (magnetization) is induced in the direction parallel to the magnetic (electric) field irrespective of the field direction.  

Meanwhile, for the magnetic toroidal dipoles, the antisymmetric transverse ME responses are activated as 
\begin{align}
\label{eq:ME_toroidal}
\mathbf{P} \propto \mathbf{T} \times \mathbf{B}, \ \ \ 
\mathbf{M} \propto  -\mathbf{T} \times \mathbf{E}.  
\end{align}
In this case, the electric polarization (magnetization) is induced in the direction perpendicular to both the toroidal dipole moment and the magnetic (electric) field.

For the magnetic quadrupoles, the symmetric transverse ME responses are activated as 
\begin{align}
\label{eq:ME_quadrupole}
P_\mu \propto M_{\mu\nu}  B_\nu, \ \ \ 
M_\mu \propto M_{\mu\nu}  E_\nu. 
\end{align}
Thus, the longitudinal traceless responses are obtained for $M_u$ and $M_v$, while the symmetric transverse responses appear for $M_{yz}$, $M_{zx}$, and $M_{xy}$. 

In order to discuss the linear ME effects for the microscopic model Hamiltonian $\mathcal{H}^c_0+\mathcal{H}^t_0+\mathcal{H}_1$, we compute the electromagnetic tensor between magnetic moments and electric currents for each ordered state by using the linear response theory as 
\begin{align}
\label{eq:ME_linear}
K_{\mu\nu} = \frac{2\pi}{i V_0} \sum_{mn \mathbf{k}} \frac{f(\varepsilon_{n\mathbf{k}})-f(\varepsilon_{m\mathbf{k}})}{\varepsilon_{n\mathbf{k}}-\varepsilon_{m\mathbf{k}}} \frac{m^{nm}_{\mu \mathbf{k}} J^{mn}_{\nu \mathbf{k}}}{\varepsilon_{n\mathbf{k}}-\varepsilon_{m\mathbf{k}}+i \delta}, 
\end{align}
where $V_0$ is the system volume, $f(\varepsilon)$ is the Fermi distribution function, and $\varepsilon_{m \mathbf{k}}$ is the eigenvalue of $\mathcal{H}=\mathcal{H}^c_0+\mathcal{H}^t_0+\mathcal{H}_1$. 
$m^{nm}_{\mu \mathbf{k}}= \langle n \mathbf{k} |  \sigma_\mu | m \mathbf{k} \rangle$ and $J^{mn}_{\nu \mathbf{k}}= \langle m \mathbf{k} | J_\nu | n \mathbf{k} \rangle = \langle m \mathbf{k} | \partial \mathcal{H}/\partial k_\nu | n \mathbf{k} \rangle$ are the matrix elements of the spin and current operators, respectively, where $| m \mathbf{k} \rangle$ is the eigenstate of the Hamiltonian corresponding to the eigenvalue $\varepsilon_{m \mathbf{k}}$. 
$K_{\mu\nu}$ represents the coefficient for the uniform magnetization with the $\mu$ component induced by the electric field in the $\nu$ direction. 
In Eq.~(\ref{eq:ME_linear}), we set $g \mu_{\rm B} e/(2h)=1$ ($g$ is the $g$-factor, $e$ is the elementary charge, and $h$ is the Planck constant).  

Note that the uniform ASOCs proportional to $\tau_0$ or $\tau_y$ play an important role in the linear ME effects. 
This is because the current operator $J_\nu = \partial \mathcal{H}/\partial k_\nu$ includes the $\tau_0$ component from the term in $\mathcal{H}$ proportional to $t_0$ and the $\tau_y$ component from the term proportional to $t_1$ and $\delta t_1$, both of which give nonzero matrix elements for $ \langle m \mathbf{k} | J_\nu | n \mathbf{k} \rangle$. 

\subsection{Cubic symmetry}
\label{sec:Cubic Symmetry}

\begin{table*}[htb!]
\begin{center}
\caption{
Linear ME responses in the staggered ordered states under the cubic symmetry. 
$M_{\mu}^{\perp}$ and $M'^{\perp}_{\mu}$ ($M_{\mu}^{\parallel}$) denote the induced magnetizations perpendicular (parallel) to the direction of an applied electric field; the subscript $\mu=x,y,z$ represents the component of the induced magnetization, and $M_{\mu}^{\perp}$ and $M'^{\perp}_{\mu}$ have different magnitudes. 
Only the order parameters leading to nonzero linear ME responses are shown. 
$\sigma_{xy}$ and $\sigma_{d}$ denote magnetic moments along the [110] and [111] directions, respectively. 
}
\label{tab:cubicME}
\scalebox{1.0}{
 \begin{tabular}{cccccc}
 \hline\hline
order & induced odd-parity multipoles & $\mathbf{E}\parallel x$& $\mathbf{E}\parallel y$& $\mathbf{E}\parallel z$ & remark \\ 
\hline
$\sigma_x$& $M_{yz}$        &---&$M^{\perp}_z$&$ M^{\perp}_y$ & $ M^{\perp}_y= M^{\perp}_z$  \\ 
$\sigma_y$&$M_{zx}$     &$M^{\perp}_z$&---&$M^{\perp}_x$ & $ M^{\perp}_x= M^{\perp}_z$  \\ 
$\sigma_z$&$M_{xy}$     &$M^{\perp}_y$&$M^{\perp}_x$&--- & $ M^{\perp}_x= M^{\perp}_y$\\ 
\hline
$\sigma_{xy}$&$(M_{yz}, M_{zx})+(T_{x}, T_{y})$     &$M^{\perp}_z$&$M^{\perp}_z$&$M'^{\perp}_x,M'^{\perp}_y$ & $M'^{\perp}_x=M'^{\perp}_y$\\ 
$\sigma_d$&\ \ \ $(M_{xy},M_{yz},M_{zx})+M_{0}$   \ \ \   &\ \ \ $M^{\parallel}_x, M^{\perp}_y, M^{\perp}_z$\ \ \ &\ \ \ $M^{\parallel}_y, M^{\perp}_x, M^{\perp}_z$ \ \ \ &\ \ \ $M^{\parallel}_z, M^{\perp}_x, M^{\perp}_y$ \ \ \ & \ \ \ $M^{\parallel}_x=M^{\parallel}_y=M^{\parallel}_z$, $M^{\perp}_x=M^{\perp}_y=M^{\perp}_z$\\ 
\hline
$\tau_y$& $M_{0}$     &$M^{\parallel}_x$&$M^{\parallel}_y$&$M^{\parallel}_z$ & $M^{\parallel}_x=M^{\parallel}_y=M^{\parallel}_z$\\ 
\hline\hline
\end{tabular}
}
\end{center}
\end{table*}

First, we discuss what type of the linear ME effect is induced by staggered orders in the cubic case. 
As discussed in Sec.~\ref{sec:Odd-parity multipoles}, the spin orders ($\mathcal{H}_1 \propto \sigma_\mu \tau_0$ where $\mu=x,y,z$) 
activate the magnetic quadrupoles $M_{yz}$, $M_{zx}$, and $M_{xy}$.
The magnetic quadrupoles are expected to give rise to the symmetric transverse ME responses in Eq.~(\ref{eq:ME_quadrupole}) 
as discussed above. 
For example, the spin order with $\langle \sigma_x \rangle \neq 0$ activates the magnetic quadrupole $M_{yz}$, which leads to the transverse magnetization along the $z$ ($y$) direction induced by the electric field in the $y$ ($z$) direction. 
In a similar manner, the transverse ME responses are expected for $\sigma_y$ and $\sigma_z$ orders, since they are accompanied with $M_{zx}$ and $M_{xy}$, respectively. 

The results can be extended to the spin orders whose magnetic moments deviate from the $\langle100\rangle$ direction. 
For example, when the magnetic moments are along the [110] direction, the order parameter $\langle \sigma_{xy} \rangle$ is represented by $\langle \sigma_x \rangle$ and $\langle \sigma_y \rangle$ as $\langle \sigma_{xy} \rangle = \sqrt{\langle\sigma_x \rangle^2 + \langle\sigma_y\rangle^2}$, where $\langle \sigma_x \rangle =\langle \sigma_y \rangle$. 
This type of order parameter leads to the toroidal dipoles $T_x$ and  $T_y$ in addition to the magnetic quadrupoles $M_{yz}$ and $M_{zx}$, since the nonzero $\langle \sigma_{xy} \rangle$ reduces the $T_{d}$ symmetry to the subgroup, and $T_{1}$ and $T_{2}$ turn to be in the same irreducible representation~\cite{Shiina_1997JPSJ_66.1741S}.
Thus, the induced toroidal dipoles are expected to become large when the effect of the symmetry lowering becomes prominent by increasing $\langle \sigma_{xy} \rangle$. 
Thus, the total ME response is obtained by the sum of the contributions from $(M_{yz}, M_{zx})$ and $(T_x, T_y)$. 
As the ME response of the toroidal dipole is the antisymmetric transverse one in Eq.~(\ref{eq:ME_toroidal}), it is expected that the magnetization induced along the $z$ direction by the electric field along the $x$ ($y$) direction is different from that along the $x$ ($y$) direction by the electric field along the $z$ direction. 

When the magnetic moments are aligned along the $\langle 111 \rangle$ direction [$\mathcal{H}_1 \propto (\sigma_x + \sigma_y +\sigma_z) \tau_0$], the order parameter $\langle \sigma_d  \rangle$ is represented by $\langle \sigma_d  \rangle = \sqrt{\langle \sigma_x \rangle^2+\langle \sigma_y \rangle^2+\langle \sigma_z \rangle^2}$, where $\langle \sigma_x \rangle =\langle \sigma_y \rangle= \langle \sigma_z \rangle$. 
In this case, the symmetry lowering due to $\langle \sigma_{d}\rangle$ reduces $A_{2}$ and $T_{1}$ into the same irreducible representation, which induces the magnetic monopole $M_{0}$ in addition to the magnetic quadrupoles $(M_{yz}, M_{zx}, M_{xy})$; see Table~\ref{tab:cubic}.
Thus, both longitudinal and transverse ME responses are expected for the $\langle \sigma_d  \rangle$ order, as the magnetic monopole results in the isotropic longitudinal ME in Eq.~(\ref{eq:ME_monopole}). 
As in the $\langle \sigma_{xy} \rangle$ case, the induced longitudinal response is expected to become larger when the symmetry lowering becomes prominent by increasing $\langle \sigma_{d}\rangle$.

Next, we discuss the ME effect in the octupole ordered state ($\mathcal{H}_1 \propto \sigma_0 \tau_y$). 
The octupole order $\langle \tau_y\rangle$ is accompanied with the magnetic monopole $M_{0}$, as shown in Table~\ref{tab:cubic}. 
Thus, from Eq.~(\ref{eq:ME_monopole}), the isotropic longitudinal ME effect is expected, as the $\langle \sigma_d \rangle$ order.

In the spin-orbital ordered states ($\mathcal{H}_1 \propto \sigma_\mu \tau_\nu$ where $\mu=x,y,z$ and $\nu=x,z$), the magnetic quadrupoles $(M_{yz}, M_{zx}, M_{xy})$ and the toroidal dipoles $(T_x, T_y)$ are induced for the $\sigma_\mu \tau_z$ and $\sigma_\mu  \tau_x$ orderings, respectively, as shown in Table~\ref{tab:cubic}. 
However, there are no ME responses for these cases in contrast to the spin or octupole ordered states discussed above. 
This is because the effective ASOC vanishes in the cubic case, as discussed in Eqs.~(\ref{eq:ASOC_spinxtauz})-(\ref{eq:ASOC_spinztaux}). 
The responses, however, become nonzero once a tetragonal distortion is introduced, as described in the following section. 

In short, in the cubic case, the staggered spin orders along the $\langle100\rangle$ directions, $\sigma_\mu$ with $\mu=x,y,z$, lead to the symmetric transverse ME responses, while the octupole $\tau_y$ order leads to the longitudinal one. 
When the magnetic moments deviate from $\langle100\rangle$ to $\langle 110 \rangle$ ($\langle 111 \rangle$), the asymmetric transverse (longitudinal) ME effects are also induced.  
We summarize the results in Table~\ref{tab:cubicME}. 
In the table, instead of the ME coefficients $K_{\mu\nu}$ in Eq.~(\ref{eq:ME_linear}), we show the induced magnetizations $M^{\perp}_\mu$ and $M^{\parallel}_\mu$; $M^{\perp}_\mu$ represents a transverse component with $\mu\neq \nu$, while $M^{\parallel}_{\mu}$ represents a longitudinal one with $\mu= \nu$. 
$M'^{\perp}_\mu$ denotes a transverse component with a different magnitude from $M^{\perp}_\mu$. 

\begin{figure*}[hbt!]
\begin{center}
\includegraphics[width=1.0 \hsize]{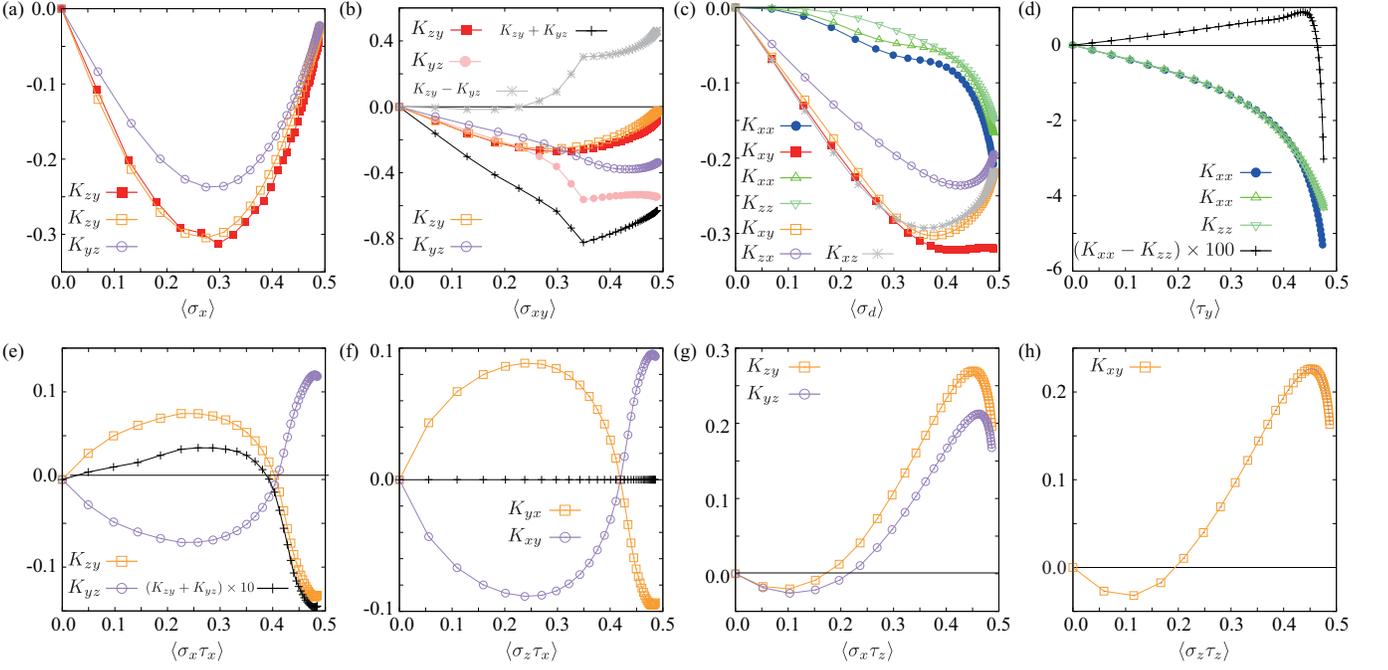} 
\caption{
\label{Fig:ME}
The ME coefficients in the staggered ordered phases with the order parameters (a) $\langle \sigma_x \rangle$, (b) $ \langle \sigma_{xy} \rangle$ $= \sqrt{\langle \sigma_x \rangle^2+\langle \sigma_y \rangle^2} $, (c) $\langle  \sigma_d \rangle$ $= \sqrt{\langle \sigma_x \rangle^2+\langle \sigma_y \rangle^2+\langle \sigma_z \rangle^2} $, (d) $\langle \tau_y \rangle$, (e) $ \langle \sigma_x \tau_x \rangle$, (f) $ \langle \sigma_z \tau_x \rangle$, (g) $ \langle \sigma_x \tau_z\rangle$, and (h) $\langle \sigma_z \tau_z \rangle$. 
The horizontal axis is the corresponding order parameter in each phase. 
The closed (open) symbols represent the results in the absence (presence) of the tetragonal crystalline electric field. 
The data are calculated at $t_0=1.0$, $t_1=0.5$, $\delta t_0 = 0$ ($\delta t_0 =0.1$), $\delta t_1 = 0$ ($\delta t_1=0.05$), $\Delta_t =0$ ($\Delta_t =0.5$), temperature $T=0.01$, and the damping factor $\delta = 0.01$ by using Eq.~(\ref{eq:ME_linear}) for the cubic (tetragonal) system. 
}
\end{center}
\end{figure*}

In order to discuss the ME responses quantitatively, we calculate the order parameter dependence of the ME coefficients in Eq.~(\ref{eq:ME_linear}). 
The results are shown by the closed symbols in Figs.~\ref{Fig:ME}(a)-\ref{Fig:ME}(d). 
Figure~\ref{Fig:ME}(a) shows $K_{zy}$ in the spin-$\sigma_x$ ordered state. 
When the order parameter is small, the ME coefficient grows linearly to the order parameter due to $\mathcal{H}^{\sigma_x}_{\rm ASOC} \propto  h$ in Eq.~(\ref{eq:ASOC_spinx}), while it approaches zero as the order parameter reaches the saturated value $0.5$. 
The result indicates that the magnitude of the transverse ME response is maximized when the ordered moment is about a half of the saturation. 
Similar results are obtained for the spin-$\sigma_y$ and $\sigma_z$ orders. 

Figure~\ref{Fig:ME}(b) shows the result for the spin-$\sigma_{xy}$ order with the polarized moments along the [110] direction. 
In this case, the transverse ME responses $K_{zy}$ ($K_{zx}$) and $K_{yz}$ ($K_{xz}$) are no longer symmetric with each other, which indicates the emergence of the additional toroidal multipole $T_x$ ($T_y$). 
The contribution from the magnetic quadrupole $K_{zy}+K_{yz}$ is dominant for the small order parameter. 
While increasing $h$, the contribution from the toroidal dipole $K_{zy}-K_{yz}$ becomes relevant, as the symmetry lowering becomes prominent by increasing $\langle \sigma_{xy} \rangle$ as mentioned above. 

Figure~\ref{Fig:ME}(c) shows the result for the spin-$\sigma_d$ order with the polarized moments along the [111] direction. 
In this case, the longitudinal ME response with $K_{xx}$ becomes nonzero in addition to the transverse one with $K_{xy}$, as shown in the figure. 
This additional $K_{xx}$ is proportional to $\langle\sigma_d\rangle^3$ for the small order parameter.
The magnitude of $K_{xx}$ is much smaller than $K_{xy}$ when the order parameter is small, e.g., the ratio of $M^{\parallel}_x$ for $M^{\perp}_x$ is about 5\% at $\langle \sigma_d \rangle \sim 0.05$. 
While increasing $\langle \sigma_d \rangle$, the difference becomes smaller, and the ratio approaches about 65\% at $\langle \sigma_d \rangle = 0.5$. 
This result indicates that the contribution from the magnetic quadrupole is dominant when the magnetic moment is small, while that from the magnetic monopole becomes relevant gradually with the growth of the magnetic moment, as expected from the arguments above. 

Figure~\ref{Fig:ME}(d) shows the ME response in the octupole ordered state with the order parameter $\langle \tau_y \rangle$. 
In this case also, we have a nonzero longitudinal response with $K_{xx}$, whose magnitude increases linearly to $\langle \tau_y \rangle$ because the induced ASOC is proportional to $h$ in Eq.~(\ref{eq:ASOC_tauy}). 
The ME coefficient is monotonically enhanced while further increasing $\langle \tau_y \rangle$.

\subsection{Tetragonal symmetry}
\label{sec:Tetragonal Symmetry}

\begin{table*}[htb!]
\begin{center}
\caption{
Linear ME responses in the staggered ordered states under the tetragonal symmetry. 
$M^{\perp}_{\mu}$, $M'^{\perp}_{\mu}$, and $M''^{\perp}_{\mu}$ ($M^{\parallel}_{\mu}$ and $M'^{\parallel}_{\mu}$) denote the induced magnetizations with different magnitudes perpendicular (parallel) to the direction of an applied electric field; the subscript $\mu=x,y,z$ represents the component of the induced magnetization. 
$\sigma_{xy}$ and $\sigma_{d}$ denote magnetic moments along the [110] and [111] directions, respectively. 
}
\label{tab:tetraME}
\scalebox{1.0}{
 \begin{tabular}{cccccc}
 \hline\hline
 order & induced odd-parity multipoles & $\mathbf{E}\parallel x$& $\mathbf{E}\parallel y$& $\mathbf{E}\parallel z$  & remark \\ 
\hline
$\sigma_x$& $M_{yz}+T_x$        &---&$M^{\perp}_z $&$ M'^{\perp}_y $  \\ 
$\sigma_y$&$M_{zx}+T_y$     &$M^{\perp}_z $&---&$M'^{\perp}_x $  \\ 
$\sigma_z$&$M_{xy}$     &$M^{\perp}_y$&$M^{\perp}_x$&--- & $M^{\perp}_x=M^{\perp}_y$\\ 
\hline
$\sigma_{xy}$&$(M_{yz}, M_{zx})+(T_{yz}, T_{zx})$     &$M^{\perp}_z$&$M^{\perp}_z$&$M'^{\perp}_x, M'^{\perp}_y$ & $M'^{\perp}_x = M'^{\perp}_y$
\\ 
$\sigma_d$&\ \ \ $(M_{yz},M_{zx})+(T_x,T_y)$ \ \ \      & \ \ \ $M^{\parallel}_x, M^{\perp}_y, M'^{\perp}_z$\ \ \  &\ \ \ $M^{\parallel}_y, M^{\perp}_x, M'^{\perp}_z $ \ \ \ &\ \ \ $M'^{\parallel}_z, M''^{\perp}_x, M''^{\perp}_y$ \ \ \   &\ \ \ $M^{\parallel}_x=M^{\parallel}_y$, $M^{\perp}_x=M^{\perp}_y$, \\ &$+M_{xy}+M_{0}+M_{u}$&&&& $M''^{\perp}_x=M''^{\perp}_y$
\vspace{1mm}
 \\ 
 \hline
$\tau_y$& $M_{0}  + M_{u}$      &$M^{\parallel}_x$&$M^{\parallel}_y$&$M'^{\parallel}_z$& $M^{\parallel}_x=M^{\parallel}_y$\\ 
\hline
$\sigma_x \tau_x$&$M_{yz}+T_x$        &---&$M^{\perp}_z$&$ M'^{\perp}_y$ \\ 
$\sigma_y \tau_x$&$M_{zx}+T_y$     &$M^{\perp}_z$&---&$M'^{\perp}_x$ \\ 
$\sigma_z \tau_x$&$T_z$     &$M^{\perp}_y$&$M'^{\perp}_x$&---&$M'^{\perp}_x=- M^{\perp}_y$ \\ 
\hline
$\sigma_x \tau_z$&$M_{yz}+T_x$        &---&$M^{\perp}_z$&$ M'^{\perp}_y$ \\ 
$\sigma_y \tau_z$&$M_{zx}+T_y$      &$M^{\perp}_z$&---&$M'^{\perp}_x$ \\ 
$\sigma_z \tau_z$&$M_{xy}$     &$M^{\perp}_y$&$M^{\perp}_x$&--- & $M^{\perp}_x=M^{\perp}_y$ \\ \hline\hline
\end{tabular}
}
\end{center}
\end{table*}

We turn to the ME effects in the tetragonal case. 
In this case, the symmetry is lowered from $T_d$ to $D_{2d}$. 
$D_{2d}$ is a subgroup of $T_d$, and the reduction rule for the irreducible representations between them is given in Table~\ref{tab:cubic}. 
Accordingly, compared to the cubic case, different odd-parity multipoles are additionally induced for some of the staggered orders, which lead to additional ME responses.  

In the spin ordered states ($\sigma_\mu \tau_0$ where $\mu=x,y,z$), the toroidal dipole $T_x$ ($T_y$) is additionally activated besides the magnetic quadrupole $M_{yz}$ ($M_{zx}$) in the $\langle \sigma_x \rangle$ ($\langle \sigma_y \rangle$) order. 
This is because the ($\sigma_x, \sigma_y$) orders belong to the irreducible representation $E$ under the $D_{2d}$ symmetry (see Table~\ref{tab:cubic}). 
Thus, the ME responses in the $\langle \sigma_x \rangle$ and $\langle \sigma_y \rangle$ orders are similar to that in the $\langle \sigma_{xy} \rangle$ order under the cubic symmetry. 
Meanwhile, as there is no additional multipole in the $\langle \sigma_z \rangle$ order which has the $A_{2}$ symmetry, it leads to the same ME responses as in the cubic case. 

In the case of the spin-$\sigma_{d}$ order, a further additional multipole $M_{u}$ is induced under the tetragonal symmetry because $M_{u}$ belongs to the same irreducible representation $B_{1}$ as $M_{0}$ (see Table~\ref{tab:cubic}). 
Hence, $(M_{xy}, M_{yz}, M_{zx})$, $(T_x, T_y, T_z)$, $M_{0}$, and $M_{u}$ contribute the ME responses in the $\sigma_{d}$ ordered state. 
Similar discussion is applied to the octupole order $\langle \tau_y \rangle $ which belongs to the irreducible representation $B_{1}$. 

In the spin-orbital ordered states ($\mathcal{H}_1 \propto \sigma_\mu \tau_\nu$ where $\mu=x,y,z$ and $\nu=x,z$), where no ME effect is expected in the cubic case, the lowering of the symmetry to tetragonal $D_{2d}$ leads to nonzero ME responses due to the emergent ASOC in Eqs.~(\ref{eq:ASOC_spinxtauz})-(\ref{eq:ASOC_spinztaux}). 
The $\sigma_z \tau_z$ order induces the magnetic quadrupole $M_{xy}$, which leads to the symmetric ME response, while the $\sigma_z \tau_x $ order induces the toroidal dipole $T_z$, which leads to the antisymmetric ME response. 
On the other hand, in the $\sigma_x \tau_\nu$ ($\sigma_y \tau_\nu$) ordered states ($\nu=x,z$), both $M_{yz}$ ($M_{zx}$) and $T_x$ ($T_y$) are activated because they belong to the same irreducible representation $E$. 
In other words,  $M_{yz}$ ($M_{zx}$) and $T_x$ ($T_y$) are indistinguishable from the symmetry point of view. 
In order to clarify which multipoles play a dominant role, it is necessary to evaluate the values of the ME coefficients in Eq.~(\ref{eq:ME_linear}); we will return to this point later. 

In short, in the tetragonal case, the staggered spin orders along the $\langle100\rangle$ directions, $\sigma_\mu$ with $\mu=x,y$, lead to the transverse ME responses, whose magnitudes depend on the electric-field direction, while there is no qualitative change in the $\sigma_z$ order from the cubic case. 
The octupole $\tau_y$ order induces the longitudinal ME responses with different magnitudes between the $(x,y)$ and $z$ components. 
In the spin-orbital channel, the $\sigma_z \tau_x$ order leads to the asymmetric ME response, while the $\sigma_z \tau_z$ order leads to the symmetric ME response. 
The $\sigma_x \tau_x$ and $\sigma_x \tau_z$ ($\sigma_y \tau_x$ and $\sigma_y \tau_z$) orders show similar ME responses to those in the $\sigma_x$ ($\sigma_y$) order. 
We summarize the results in Table~\ref{tab:tetraME}. 
The notations are similar to those in Table~\ref{tab:cubicME}. 

As in the cubic case, we calculate the ME coefficients in Eq.~(\ref{eq:ME_linear}) as functions of the order parameters under the tetragonal symmetry. 
The results are presented by the open symbols in Figs.~\ref{Fig:ME}(a)-\ref{Fig:ME}(h).
In the $\langle \sigma_x \rangle$ ordered state, $K_{zy}$ behaves differently from $K_{yz}$, as shown in Fig.~\ref{Fig:ME}(a). 
This is due to the emergence of the additional toroidal multipole $T_x$. 
We note that $K_{yz}$ changes from the cubic case more significantly compared to $K_{zy}$. 
This is because the contribution from the first term in Eq.~(\ref{eq:ASOC_spinx}) is larger than that from the second term in the additional ASOC induced by the tetragonal distortion for $\delta t_1>0$. 
For other spin and orbital orders, we find quantitative changes of the ME responses; the results for $\sigma_{xy}$, $\sigma_d$, and $\tau_y$ are shown in Figs.~\ref{Fig:ME}(b), \ref{Fig:ME}(c), and \ref{Fig:ME}(d), respectively.

Figures~\ref{Fig:ME}(e)-\ref{Fig:ME}(h) show the results for the spin-orbital orders with the order parameters $\langle \sigma_\mu \tau_\nu \rangle$ ($\mu=x,y,z$ and $\nu=x,z$). 
In the cases of $\sigma_z \tau_x$ [Fig.~\ref{Fig:ME}(f)], $K_{yx}$ and $K_{xy}$ become nonzero with the same magnitude but different sign. 
This is attributed to the presence of the toroidal dipole $T_z$, as discussed above. 
The toroidal-type ME responses are enhanced when the ordered parameter is about a half and full of 
the saturation. 
On the other hand, in the $\sigma_x \tau_x$ ordered state [Fig.~\ref{Fig:ME}(e)], $K_{zy}$ has a different magnitude from $K_{yz}$. 
This is because there is a contribution from the magnetic quadrupole $M_{yz}$ in addition to $T_x$ in the $\sigma_x \tau_x$ ordered state. 
Nevertheless, we find that the magnitude of $K_{zy}$ is close to that of $K_{yz}$ in Fig.~\ref{Fig:ME}(e), which indicates that the contribution from $T_x$ is much larger than $M_{yz}$. 
In a similar way, we conclude that the contribution from the magnetic quadrupole is larger than the magnetic toroidal dipole in the $\sigma_x \tau_z$ ordered state in Fig.~\ref{Fig:ME}(g) by comparing the ME response in the $\sigma_x \tau_x$ ordered state in Fig.~\ref{Fig:ME}(h).

It is also interesting to point out the sign change of the ME responses in Figs.~\ref{Fig:ME}(e)-\ref{Fig:ME}(h), which might enable us to control the direction of the induced magnetization by external parameters, such as temperature and pressure. 
The origin of such a sign change will be left for future problem. 

\section{Summary and concluding remarks}
\label{sec:Summary}

In summary, we have investigated the linear ME effects associated with odd-parity multipoles, with $5d$ transition metal oxides $A$OsO$_4$ ($A=$ K, Rb, and Cs) in mind. 
By taking into account the mixing between $e_g$ and $t_{2g}$ orbitals by the SOC, we constructed an effective two-orbital model for the $e_g$ orbital manifold. 
We have classified the staggered spin and orbital orders and associated odd-parity multipoles, such as magnetic monopole, quadrupoles, and toroidal dipoles, from the symmetry point of view. 
In each electronic ordered state, we have shown the explicit form of the effective ASOC generated by the staggered order. 
We have also predicted what type of the linear ME effect is expected in the system, which provides a reference to identify the order parameter for $A$OsO$_4$ by measurement. 

Let us comment on yet another interesting aspect by the emergent ASOCs and odd-parity multipoles as the future problem. 
As discussed in Sec.~\ref{sec:Introduction}, the emergent ASOCs result in asymmetry in the electronic structure and spin-wave excitations. 
Such asymmetry can be a source of nonreciprocal phenomena in optical and magnetic collective excitations. 
In the present model, such nonreciprocal responses are expected for particular staggered electronic orders. 
For example, in the spin-orbital $\sigma_x \tau_x$ ordered state, the effective ASOC in Eq.~(\ref{eq:ASOC_spinxtaux}) leads to a band deformation with the band bottom shift, while in the spin $\sigma_x$ ordered state, the valley splitting in the band structure is caused by the $k^3$ contribution from the ASOC in Eq.~(\ref{eq:ASOC_spinx}).  
It is intriguing to examine how these modulated band structures are related with the nonreciprocal phenomena, which will provide further probes to detect odd-parity multipoles.

Our results offer a way to observe odd-parity multipoles, such as magnetic quadrupoles and toroidal dipoles, on the diamond structure. 
There are other candidate materials with the diamond structure that possess such odd-parity multipoles. 
For example, KRuO$_4$, whose crystal structure is similar to $A$OsO$_4$, shows a staggered magnetic order along the $z$ direction~\cite{marjerrison2016structure,xu2017negligible}, which implies the emergence of magnetic quadrupoles. 
Spinels $AB$$_2$O$_4$ are another candidates, as the $A$ sites comprise the diamond structure. 
Further experiments are desired for these systems on the linear ME effects as well as the nonreciprocal optical responses.

\begin{acknowledgments}
We thank Z. Hiroi and J. Yamaura for fruitful discussions. 
This research was supported by JSPJ KAKENHI Grants Numbers 15K05176, 15H05885 (J-Physics), and 16H06590. Parts of the numerical calculations were performed in the supercomputing systems in ISSP, the University of Tokyo.
\end{acknowledgments}

\bibliographystyle{apsrev}
\bibliography{ref}

\end{document}